\documentclass[preprint2]{aastex}

\usepackage{natbib}
\newcommand{\mdot}{M$_{\odot}$~}

\shorttitle{Is the observed high-frequency radio luminosity distribution of QSOs bimodal?}
\shortauthors{Mahony et al.}

\begin{document}


\title{Is the observed high-frequency radio luminosity distribution of QSOs bimodal?}


\author{Elizabeth K. Mahony\altaffilmark{1}, Elaine M. Sadler, Scott M. Croom}
\affil{Sydney Institute for Astronomy, School of Physics, The University of Sydney, NSW 2006, Australia}
\email{emahony@physics.usyd.edu.au}

\author{Ronald D. Ekers, Ilana J. Feain}
\affil{Australia Telescope National Facility, CSIRO Astronomy and Space Science, P.O Box 76, Epping, NSW 1710, Australia}

\author{and Tara Murphy}
\affil{Sydney Institute for Astronomy, School of Physics, The University of Sydney, NSW 2006, Australia}
\affil{School of Information Technologies, The University of Sydney, NSW 2006, Australia}


\altaffiltext{1}{Present address: ASTRON, the Netherlands Institute for Radio Astronomy, Postbus 2, 7990 AA, Dwingeloo, The Netherlands.}


\begin{abstract}

The distribution of QSO radio luminosities has long been debated in the literature. Some argue that it is a bimodal distribution, implying that there are two separate QSO populations (normally referred to as `radio-loud' and `radio-quiet'), while others claim it forms a more continuous distribution characteristic of a single population. We use deep observations at 20\,GHz to investigate whether the distribution is bimodal at high radio frequencies. Carrying out this study at high radio frequencies has an advantage over previous studies as the radio emission comes predominantly from the core of the AGN, hence probes the most recent activity. Studies carried out at lower frequencies are dominated by the large scale lobes where the emission is built up over longer timescales ($10^7-10^8$ yrs), thereby confusing the sample. Our sample comprises 874 X-ray selected QSOs that were observed as part of the 6dF Galaxy Survey. Of these, 40\% were detected down to a 3$\sigma$ detection limit of 0.2--0.5\,mJy. 

No evidence of bimodality is seen in either the 20\,GHz luminosity distribution or in the distribution of the R$_{20}$ parameter: the ratio of the radio to optical luminosities traditionally used to classify objects as being either radio-loud or radio-quiet. Previous results have claimed that at low radio luminosities, star formation processes can dominate the radio emission observed in QSOs. We attempt to investigate these claims by stacking the undetected sources at 20\,GHz and discuss the limitations in carrying out this analysis. However, if the radio emission was solely due to star formation processes, we calculate that this corresponds to star formation rates ranging from $\sim10$\mdot\,yr$^{-1}$ to $\sim2300$\mdot\,yr$^{-1}$. 
 
\end{abstract}





\section{Introduction}

QSOs are often classified into two broad categories based on their radio properties: radio-loud and radio-quiet, where the `radio-loudness' of a QSO is usually defined by the ratio of its radio to optical luminosity (\citealt{kellermann89} and references therein). However, the underlying distribution of QSO radio luminosities has long been debated in the literature. There are two opposing views: the first is that the distribution is bimodal with approximately 5--10\% of QSOs being radio-loud \citep{ivezic02, jiang07}, and the second is that there is a broad, continuous distribution with no clear dividing line between radio-loud and radio-quiet QSOs \citep{cirasuolo03,rafter09}.  

Clarifying this issue can provide insight into the physics associated with forming radio jets. A bimodal distribution suggests that there are two intrinsically different classes of QSOs, only one of which is able to become a strong radio source, while a continuous distribution suggests that all QSOs have low-luminosity radio sources which become stronger during episodes of unusually high activity.

Although there have been numerous studies addressing this issue, the wide range of selection criteria used to define the samples observed, along with the different flux density limits reached, contribute to a large range of contradicting results. Studies that report a bimodal distribution (i.e. two different QSO populations) include \citet{strittmatter80, kellermann89, mpm90, stocke92, hooper95} and \citet{ivezic02, ivezic04}. On the other hand, studies by \citet{condon80, white00, cirasuolo03} and \citet{rafter09} argue that the range of radio luminosities produces a more continuous distribution which includes a population of `radio-intermediate' QSOs. 

With the completion of large area surveys such as the Sloan Digital Sky Survey (SDSS; \citealt{sdss}), Faint Images of the Radio Sky at Twenty centimetres (FIRST; \citealt{first}) and the NRAO VLA Sky Survey (NVSS; \citealt{nvss}) came a shift from follow-up observations of smaller samples to crossmatching of large catalogs (e.g. \citealt{ivezic02, cirasuolo03, jiang07}). These studies have led to suggestions that the fraction of radio-loud sources observed is correlated with a range of observed properties including the optical magnitude and redshift \citep{padovani93, jiang07}, optical colors \citep{white07}, black hole mass \citep{best05}, accretion rate \citep{sikora07, shankar10, broderick11}, host galaxy morphology \citep{stawarz08} and stellar mass \citep{best05,mauch07}. 

Another reason why different studies have produced contradictory results is that many of the previous radio studies have used QSO samples which span a wide range in redshift and a relatively narrow range in optical magnitude. This introduces a strong and spurious correlation between redshift and optical luminosity, which in combination with the rapid cosmic evolution of the QSO population introduces selection effects which greatly complicate any interpretation of the data \citep{jiang07}.

By carrying out this study at high radio frequencies, we probe the radio emission much closer to the core of the AGN. As such, we can gain insight into the most recent activity, comparable to timescales seen in the optical regime. At lower radio frequencies the large scale radio lobes dominate the emission which are built up on timescales of $10^7-10^8$\,yrs and do not necessarily relate to the current AGN activity.

Upgrades of existing radio facilities to significantly increased bandwidth such as the Compact Array Broadband Backend (CABB; \citealt{cabb}) on the Australia Telescope Compact Array (ATCA) and the Expanded Very Large Array (EVLA: \citealt{evla}) has enabled studies of large numbers of objects to much deeper flux density limits, probing further into the radio-quiet regime. This has recently been shown in \citet{kimball11} who observed a sample of 179 optically-selected QSOs in the redshift range $0.2<z<0.3$ with the EVLA down to an rms of 6--8$\mu$Jy. The authors detect nearly all sources in the sample (97\%) and find that the resulting radio luminosity function is consistent with 2 radio source populations: those dominated by AGN emission and those dominated by star formation in the host galaxy.

In this paper, we observe a sample of X-ray selected QSOs with $z<1$ that were observed as part of the 6dF Galaxy Survey (6dFGS; \citealt{6df2004}) at 20\,GHz to study the bimodality of the radio luminosity distribution of X-ray selected QSOs at high radio frequencies. Section \ref{sampleselection} details the sample selection and Section \ref{observations} describes the follow-up radio observations. Results of this analysis are presented in Section \ref{results}, followed by a discussion on stacking the undetected sources in Section \ref{discussion}. We conclude in Section \ref{conclusions}. 

\section{Sample Selection} \label{sampleselection}

Previous studies addressing the dichotomy of radio-loud and radio-quiet QSOs have generally been based on optically selected samples which are then followed up at radio wavelengths. We take a slightly different approach by starting with an X-ray selected AGN sample. Targets were selected from the RASS--6dFGS catalog \citep{rass6df}; a catalog of 3405 AGN selected from the ROSAT ALL Sky Survey (RASS) Bright Source catalog \citep{rass} that were observed as part of the 6dF Galaxy Survey (6dFGS; \citealt{6df2004}). The 6dF Galaxy Survey is a spectroscopic redshift survey covering the entire southern sky with $|b|>10^{\circ}$. Bright X-ray sources were selected such that AGN activity was the primary source of X-ray emission, rather than X-ray emission from clusters. Of these, 2224 were observed as part of the 6dFGS\footnote{This X-ray selected sample was an additional target sample in the 6dF survey meaning that sources were observed with any spare fibres that were available} of which 77\% have reliable spectroscopic redshifts. For sources where a reliable spectrum was not able to be obtained, this was primarily due to low signal-to-noise ratios such that absorption features could not be identified. These sources would not be included in our QSO sample due to the lack of broad emission lines (as described below) so the 23\% of sources without reliable redshifts should not add any significant additional bias.  

To ensure the sample comprised solely QSOs, only targets with M$_{b_{\rm J}}<-22.3$ and broad emission lines in the optical spectrum were selected\footnote{This absolute magnitude limit corresponds to the standard QSO criteria of M$_{b_{\rm J}}<-23$ introduced by \citet{1983ApJ...269..352S} shifted to current cosmological parameters with H${_0}=71$\,km\,s$^{-1}$\,Mpc$^{-1}$ and $\Omega_{\rm m}=0.27$.}. This absolute magnitude cutoff ensures that the optical luminosity is dominated by nuclear emission with negligible contribution from the galaxy continuum. We also placed a redshift cut of $z=1$ which leaves us with a manageable sample, yet still allows us to investigate any trends with redshift. The final sample contains 874 objects.

\section{High-frequency Radio Observations} \label{observations} 

In contrast to previous studies which generally use radio emission at either 5 or 1.4\,GHz to determine the radio luminosity distribution, we have observed our sample at a higher frequency of 20\,GHz. Carrying out this study at 20\,GHz means that the core of the AGN is the dominant source of emission and provides insight into the most recent AGN activity. At lower radio frequencies, the emissions observed is dominated by the radio lobes which could be relics of past activity, thereby confusing the sample. 

\subsection{The AT20G Survey} 

Before reobserving the 874 targets which comprise the full sample, we crossmatched the sample with the Australia Telescope 20\,GHz survey (AT20G; \citealt{at20g}) to search for previous detections. The AT20G survey is a blind survey of the southern sky down to a flux density limit of 40\,mJy. The final catalog consists of 5890 sources and is dominated by flat spectrum sources \citep{at20ganalysis}. We found 56 sources in our X-ray selected sample that were detected in the AT20G survey and therefore were not followed up any further.

\subsection{New Observations} \label{newobservations}

The remaining 818 X-ray selected QSOs were observed with the Australia Telescope Compact Array (ATCA) from 2008--2010. The observations were carried out using the compact, hybrid configuration (H168) over four separate observing runs. These are summarised in Table \ref{obstab}.

The first observing run carried out in October 2008 used the old correlator that had a bandwidth of 128\,MHz in each of the two IFs (hereafter denoted as 2$\times$128). Central frequencies of 18752 and 21056\,MHz were chosen in order to match the observing setup of the AT20G survey. In early 2009 the new Compact Array Broadband Backend (CABB) was installed on the ATCA, increasing the bandwidth to 2$\times$2\,GHz \citep{cabb}. The remaining three observing runs used this new setup with central frequencies of 19000 and 21000\,MHz.

The observations were carried out in a two-step process; all sources were observed for 2$\times$40\,sec cuts and targets not detected in this time were then reobserved for 2$\times$5\,min. A small area of sky was affected by poor weather in all scheduled observing runs and therefore 18 sources were unable to be observed. These sources have been removed from the sample. 

\begin{table*}
\begin{center}
\caption{List of observations for this program. All the observations from 2008--2010 used the Hybrid 168m array configuration to obtain better (u,v) coverage. The October 2008 run was using the old correlator (2$\times$128\,MHz bandwidth) and all other runs used CABB (2$\times$2\,GHz bandwidth). \label{obstab}}
\begin{tabular}{lcccc}
\hline
{\bf Date of} &  {\bf Central } & {\bf Bandwidth} &{\bf Time on} & {\bf approx. 3$\sigma$} \\
{\bf Observations} & {\bf Frequencies (MHz)} & {\bf(MHz)} & {\bf Source} & {\bf Detection limit} \\
\hline
October 2008 & 18752, 21056 & 2$\times$128 & 2$\times$40s & 3\,mJy \\
April 2009 &  19000, 21000 & 2$\times$2000 & 2$\times$40s & 0.5\,mJy \\
October 2009 & 19000, 21000 & 2$\times$2000 & 2$\times$40s & 0.5\,mJy \\
 & 19000, 21000 & 2$\times$2000 & 2$\times$5m & 0.2\,mJy \\
March 2010 & 19000, 21000 & 2$\times$2000 & 2$\times$5m & 0.2\,mJy \\
\hline
\end{tabular}
\end{center}
\end{table*}

\begin{figure*}
\epsscale{1.5} 
 \centerline{\plotone{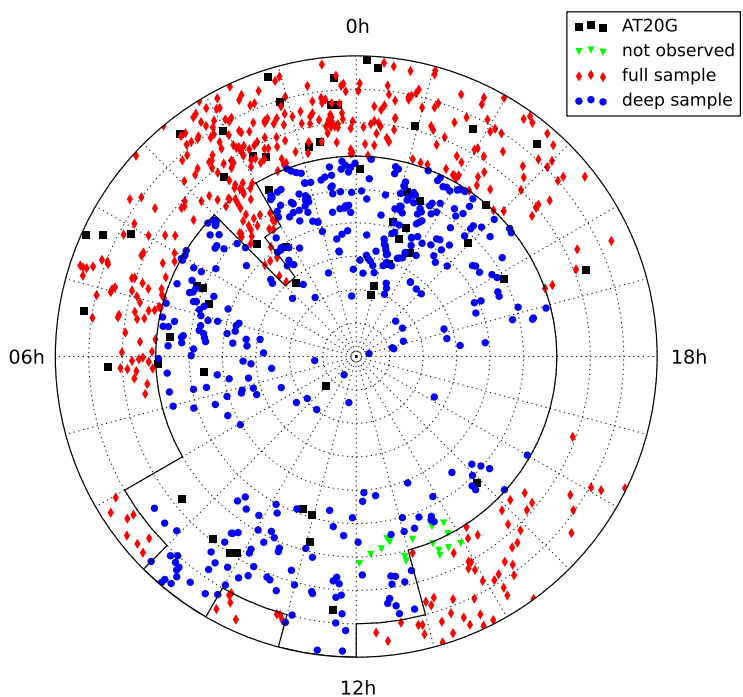}}
  \caption{Polar projection of the source distribution. The black squares indicated sources that were detected in the AT20G sample and therefore were not followed up as part of these observations. Green triangles were sources that were not observed due to poor weather and therefore removed from the sample. The blue circles are objects in the `deep sample' which contains sources that, if not detected in 2$\times$40\,s, were reobserved for an additional 2$\times$5\,m. The area of sky defining the deep sample is shown by the solid line. Red diamonds indicate those sources that were only observed for 2$\times$40\,s and are included in the `full sample'.}
\label{sourcedist}
\end{figure*}

\epsscale{1.0} 

\subsection{The deep sample}

Due to poor weather during the observations we were unable to observe all sources for the 2$\times$5\,m integration time intended. A `deep sample' was defined by a region of sky (shown in Figure \ref{sourcedist}) in which sources that were not detected in 2$\times$40\,s were reobserved for an additional 2$\times$5\,m. The deep sample covers most of the sky south of $-30^{\circ}$ declination and a few other regions which were observed during good weather. Note that the deep sample is defined only by the area of sky covered and includes sources that were detected in the initial 2$\times$40\,s observations. This ensures there is no bias by only including the fainter radio sources. There are 397 objects in the deep sample. 

\subsection{Data reduction} \label{datareduction}

All data were reduced using MIRIAD \citep{miriad}. Images were deconvolved only if a 3$\sigma$ peak was detected at the optical position of the source, or a 5$\sigma$ peak elsewhere in the field. Since we are only interested in emission from the core of the AGN, of which the position is known from the 6dF observations, fluxes and associated errors were obtained by fitting point sources at the optical position in the deconvolved images using the MIRIAD task IMFIT. As the optical position was known to better than 1$\arcsec$ accuracy, a detection at 20\,GHz was accepted if the flux density exceeded the 3$\sigma$ threshold.

In cases where other radio sources were detected in the field, these sources were modelled in the (u,v) plane and removed from the visibilities. The field was then imaged again with these sources removed. This removed any sidelobes which may affect the flux density measurement of the central source. Due to the small field of view and low source density at 20\,GHz, the majority of images had at most one other radio source in the field that needed to be removed. In a very small number of cases this process meant that radio lobes possibly associated with the X-ray/optical source were also removed. However, as we are only interested in the core emission to probe the most recent activity as opposed to emission from previous epochs of activity, this does not affect the results of this study.

\section{Results} \label{results}

\subsection{Detection rates}

\begin{figure}
\centerline{\plotone{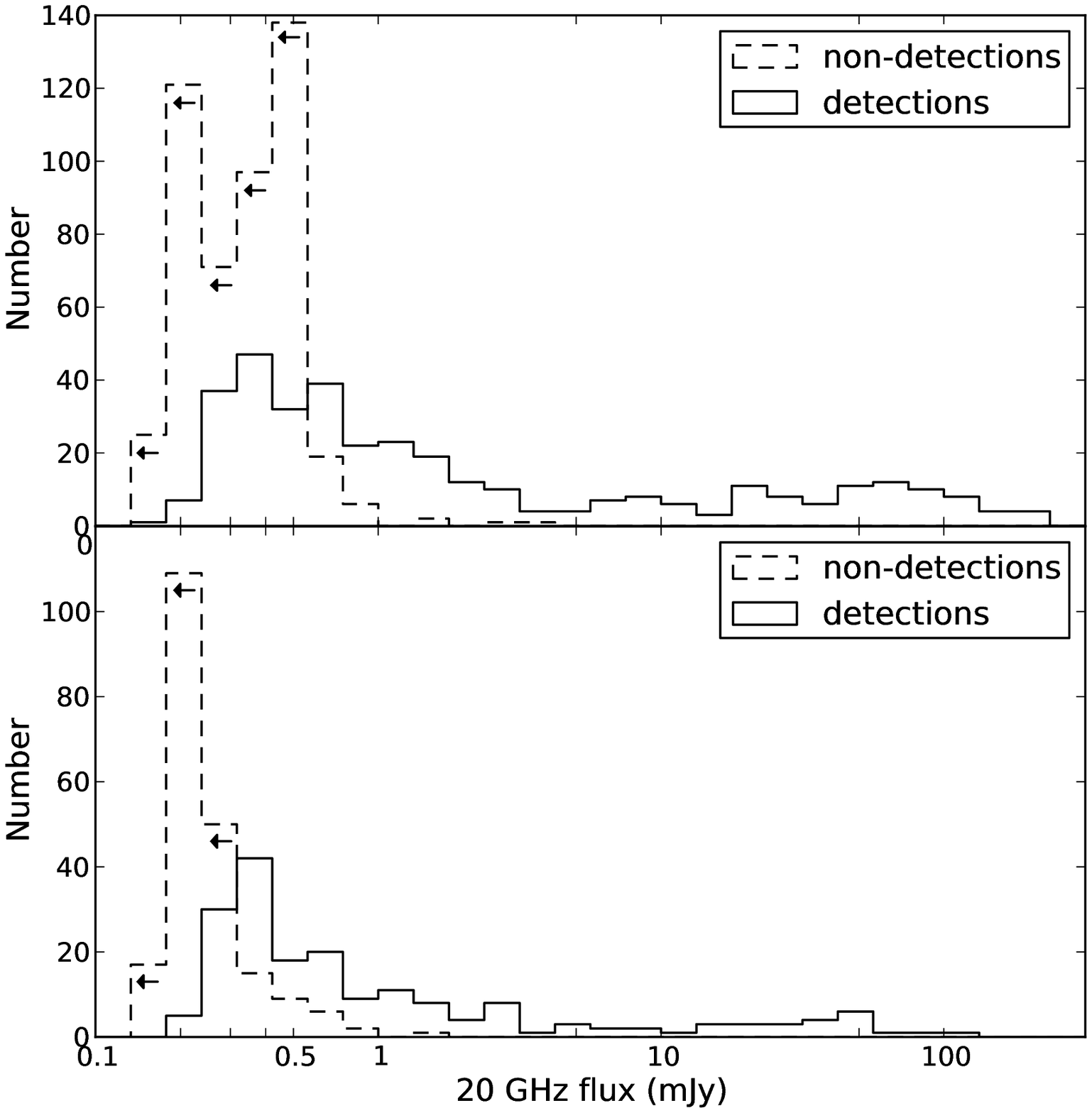}}
\caption{Flux density distribution for the full sample (top panel) and the deep sample (bottom panel), binned in log flux. Sources that were not detected are shown as upper limits by the dashed histogram and detections shown by the solid histogram.  }
\label{s20dist}
\end{figure}

Of the 800 sources observed, 321 (40\%) were detected above the 3$\sigma$ threshold. Figure \ref{s20dist} shows the flux density distribution of both the full sample and the deep sample. Detections are shown by the solid line and 3$\sigma$ upper-limits shown by the dashed line. The distribution of the upper-limits in the full sample clearly shows the effect of the different integration times for the sample. The non-detection peak at approximately 0.2\,mJy corresponds to those sources that were observed for the extra 2$\times$5 minute integrations and the peak at $\sim$0.5\,mJy corresponds to those sources that were only observed for 2$\times$40 second integrations. For the 397 objects in the deep sample, 187 (47\%) were detected. 

\begin{figure}
  \centerline{\plotone{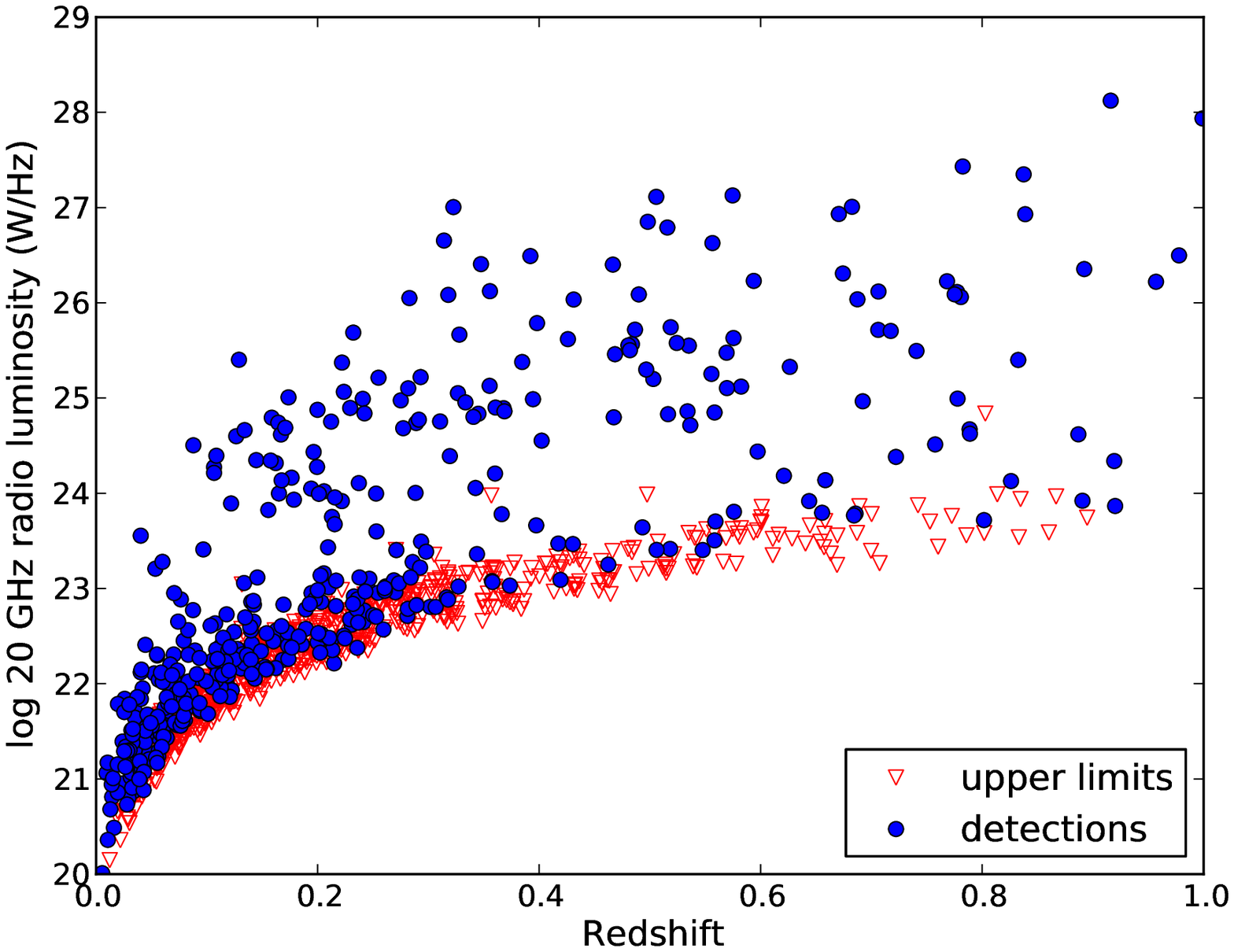}}
  \caption{20\,GHz radio luminosity for the full sample as a function of redshift. The blue circles are sources which were detected at 20\,GHz, while the red triangles indicate 3$\sigma$ upper limits.}
\label{zrlum}
\end{figure}

Redshifts were known for all sources in the sample and radio luminosities were calculated. We assumed a flat spectral index of $\alpha=0$ when applying K-corrections since we are focussing on the 20\,GHz radio emission that comes predominately from the core of the AGN. The median spectral indices of sources in the AT20G survey are $\alpha_1^5=-0.16$, $\alpha_5^8=-0.16$ and $\alpha_8^{20}=-0.28$ indicating that high frequency radio sources have flat spectral indices over a large frequency range \citep{at20ganalysis}. Assuming a spectral index of $\alpha_1^{20}=-0.5$ instead, only shifts the luminosity by 0.02 (in logspace) for a source at $z=0.1$ and 0.15 for a source at $z=1$. The 20\,GHz luminosities as a function of redshift are shown in Figure \ref{zrlum}. 

\begin{figure}
  \centerline{\plotone{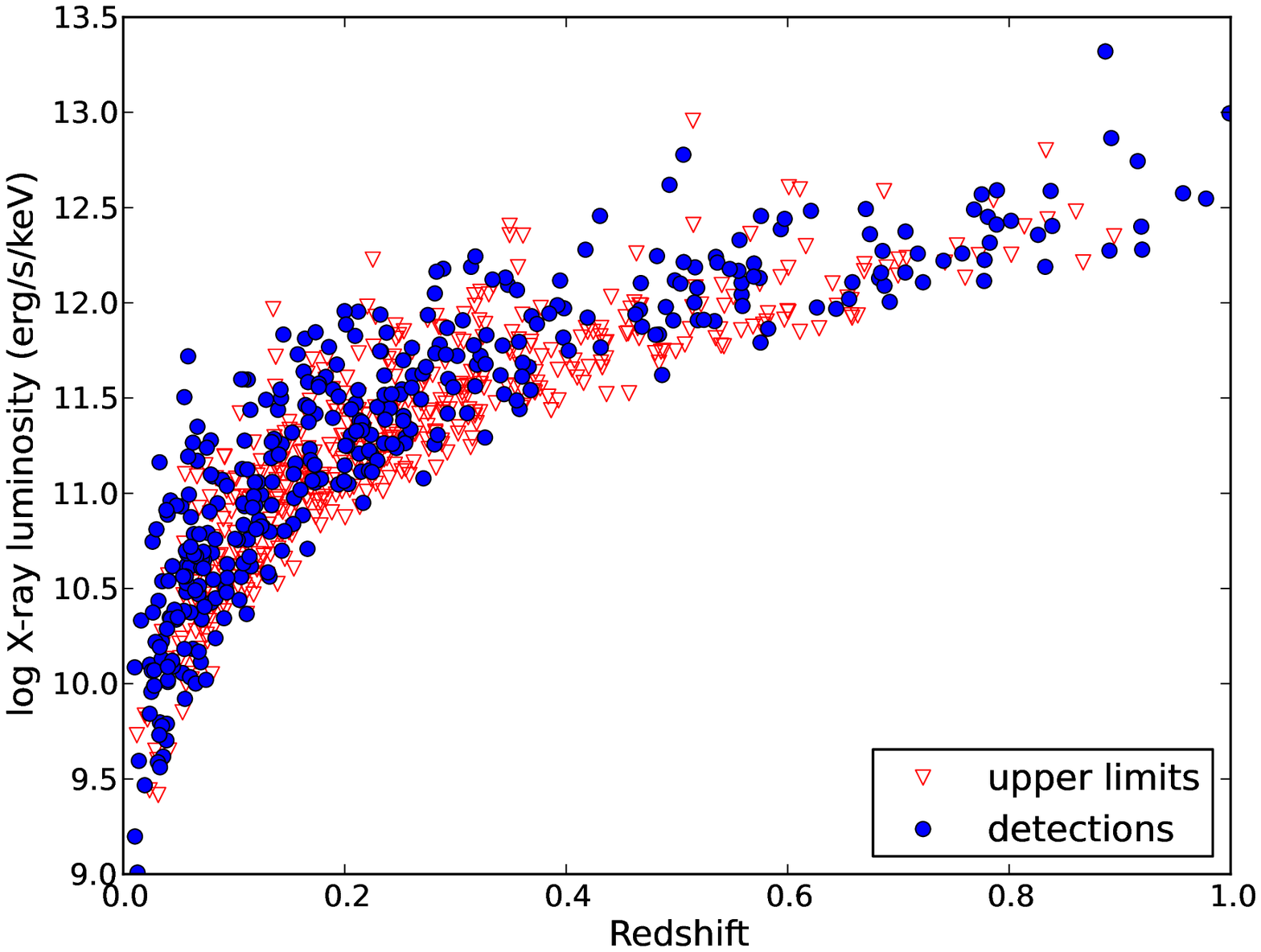}}
  \caption{X-ray luminosity against redshift for the full sample. The blue circles are sources which were detected at 20\,GHz, while the red triangles indicate 3$\sigma$ upper limits.}
\label{zxlum}
\end{figure}

\begin{figure}
  \centerline{\plotone{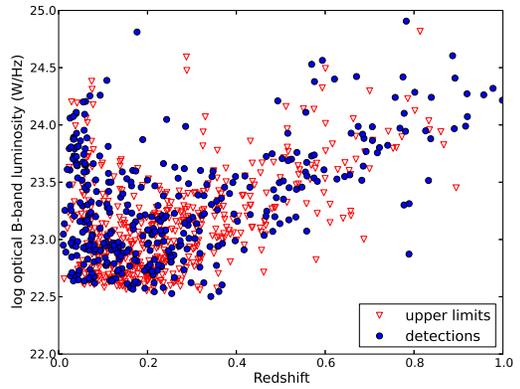}}
  \caption{Optical luminosity against redshift for the full sample. Again the blue circles are sources which were detected at 20\,GHz, while the red triangles indicate 3$\sigma$ upper limits.}
\label{zoplum}
\end{figure}

As with any flux density limited sample, this sample is subject to a luminosity-redshift degeneracy as can be seen in Figure \ref{zrlum}. To better illustrate the selection effects of the sample, Figures \ref{zxlum} and \ref{zoplum} show the X-ray and optical luminosities respectively as a function of redshift. 

To quantify the `radio-loudness' of an object we have also calculated the R$_{20}$ parameter; the ratio of the radio to optical luminosities:

\begin{equation} \label{r20}
R_{20}=\log (\frac{L_{20}}{L_{op}})
\end{equation}

The b$_{\rm J}$-band magnitudes were used to calculate the optical luminosities, and shifted into the rest-frame using a spectral index of $\alpha=-0.5$, typical for QSOs. Throughout this paper we use the following cosmological parameters: $H_0 = 71$ km s$^{-1}$ Mpc$^{-1}$, $\Omega_m=0.27$ and $\Omega_{\Lambda}=0.73$ \citep{wmap7}.

\subsection{Radio luminosity distributions} \label{luminositydistributions}

\begin{figure}
  \centerline{\plotone{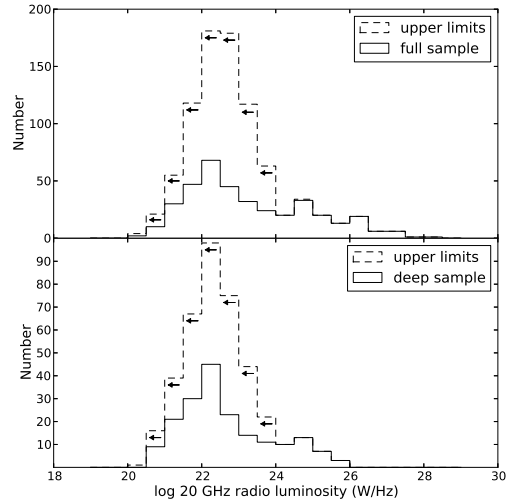}}
  \caption{Radio luminosity distribution for all sources (top panel) and the deep sample (bottom panel).}
\label{all_rlumdist}
\end{figure}

The 20\,GHz radio luminosity distributions for both the full sample (top panel) and deep sample (bottom panel) are shown in Figure \ref{all_rlumdist}. Sources not detected are shown as 3$\sigma$ upper limits by the dashed line. The ambiguity surrounding whether the luminosity distribution is bimodal or not also means that there are varying definitions for classifying an object as `radio-loud' in the literature. An often used dividing line separating radio-loud and radio-quiet sources is defined as P$_{20}=10^{24}$\,W/Hz (e.g. \citealt{mpm90, goldschmidt99})\footnote{The division at 10$^{24}$ is usually defined at frequencies of 1.4 or 5\,GHz, but using the assumption that QSOs have a flat spectrum, it will be the same at 20\,GHz}. Using this luminosity as a approximation to separate the radio-loud and radio-quiet sources means that although not all sources were detected, our observations were sensitive enough such that all of the sources that weren't detected fall into the radio-quiet regime. As such, any minimum in the distribution corresponding to the division of radio-loud and radio-quiet sources would be seen. Figure \ref{all_rlumdist} shows no indication of a bimodality. 

\begin{figure}
  \centerline{\plotone{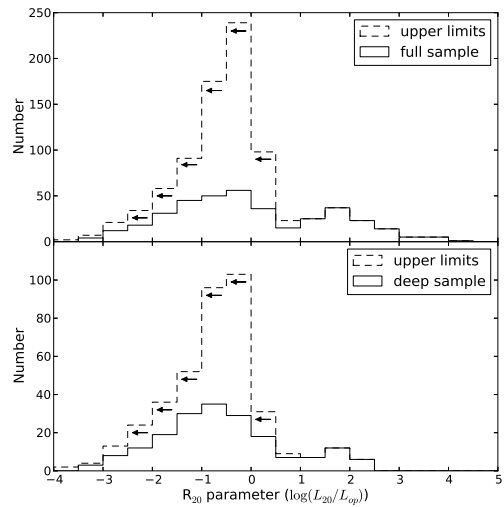}}
  \caption{R$_{20}$ distribution for all sources (top panel) and the deep sample (bottom panel). The R$_{20}$ parameter is defined as the log of the ratio of the 20\,GHz rest frame luminosity and the optical (rest frame b$_{\rm J}$-band) luminosity.}
\label{all_Rhist}
\end{figure}

The distribution of the R$_{20}$ parameter for both the full and deep samples is shown in Figure \ref{all_Rhist} with upper limits again shown by the dashed line. Here the division between radio-loud and radio-quiet sources is typically given as R$_{20}=1$ \citep{kellermann89, hooper95, jiang07}, again meaning that we have detected all the radio-loud sources. Figure \ref{all_Rhist} also shows no clear evidence for a bimodal distribution. 

\subsubsection{Selection effects}

One major limitation in compiling a flux density limited sample that covers a large range in redshift is that we are sensitive to different luminosities as a function of redshift. At high $z$ we only detect the most luminous sources while at low $z$ we observe the more common, less luminous sources, but miss the bright sources due to the small volume sampled. 

\begin{figure}
  \centerline{\plotone{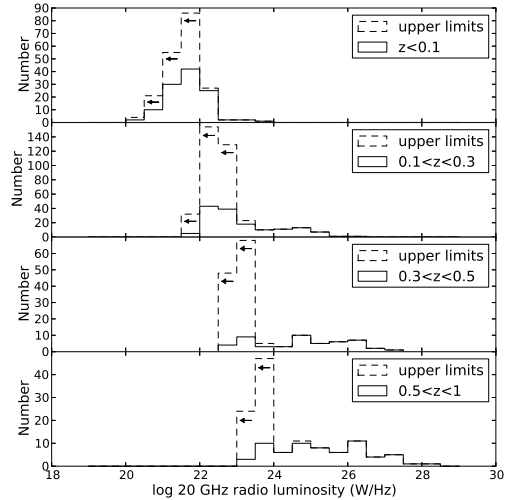}}
  \caption{20\,GHz radio luminosity distribution for the full sample separated into redshift bins. There is no evidence for a bimodal distribution in any redshift range up to $z=1$.}
\label{all_rlumhistz}
\end{figure}

\begin{figure}
  \centerline{\plotone{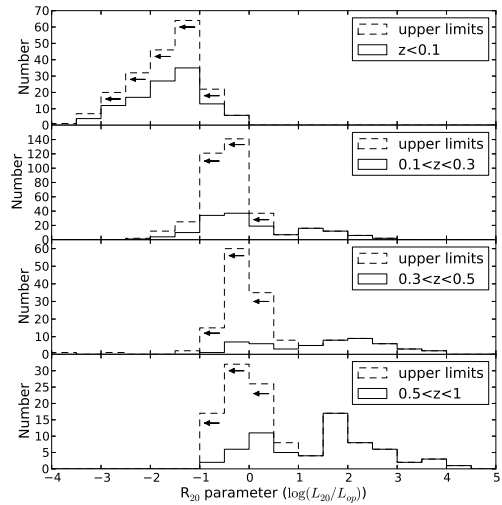}}
  \caption{R$_{20}$ distribution for the full sample separated into redshift bins.}
\label{all_Rhistz}
\end{figure}

This effect is clearly shown in Figure \ref{all_rlumhistz} which plots the 20\,GHz radio luminosity distribution split into separate redshift bins. In every redshift range there is a broad range of radio luminosities observed, but no evidence for any bimodality. Figure \ref{all_Rhistz} shows the R$_{20}$ distribution split into separate redshift bins where we see the same redshift-luminosity degeneracy. There is a clear shift from a predominately radio-quiet population at low redshifts, to a radio-loud population at high redshifts due purely to the flux density limit of the sample. 

Although we are using an X-ray selected sample, applying the condition that all objects have a 6dF spectrum adds an additional optical flux limit. This optical flux limit is made even stricter by adding the condition that the absolute magnitude must be brighter than M$_{b_{\rm J}}=-22.3$ leading to a narrow range of optical luminosities observed. While the sample presented here covers a relatively large redshift range from $0<z<1$, many previous studies place no redshift limit at all. It is this redshift-luminosity degeneracy, plus the numerous different selection criteria, that have most likely lead to contradictory results in previous studies on this subject.

\subsubsection{The validity of the R parameter}

\begin{figure}
  \centerline{\plotone{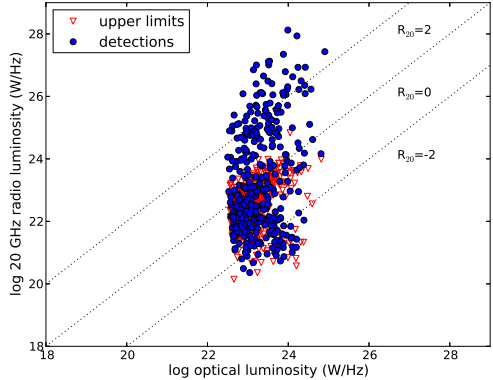}}
  \caption{20\,GHz radio luminosity against b$_{\rm J}$-band optical luminosity. The dotted lines mark lines of constant R$_{20}$. }
\label{oplumrlum}
\end{figure}

It has previously been shown that the R parameter is a meaningful measure of the radio loudness only if the optical and radio luminosities are correlated \citep{goldschmidt99}. Figure \ref{oplumrlum} shows that this is not the case. Our sample covers a wide range in radio luminosity (more than 8 orders of magnitude), but a much narrower range in optical luminosity ($\sim$2 orders of magnitude). 

However, since the R parameter is the ratio of optical to radio luminosities, it does provide useful information on the underlying physics operating in these systems. Taking the optical luminosity as a proxy for the accretion power and the radio luminosity as a proxy for the jet power, from the ratio we can gain insight into the accretion efficiency of these objects. The fact that we observe a narrow range in optical luminosities, but a very broad range of radio luminosities indicates that the accretion power does not map directly to the radio jet power observed. This suggests that other parameters (such as the magnetic field or the spin of the central black hole) can play a large role in determining the radio output observed in QSOs. 

\section{Discussion} \label{discussion}

The continuous distribution of high-frequency radio luminosities suggests that radio-loud and radio-quiet QSOs are not two intrinsically different populations. However, it has recently been suggested that while the distribution of radio luminosity of QSOs does not appear bimodal, it can be fitted by two populations: AGN and star formation \citep{kimball11}. This was based on a study of an optically selected sample of QSOs that were observed at 6\,GHz using the EVLA. 

Unfortunately the large number of non detections in our sample makes it difficult to directly compare our 20\,GHz luminosity distribution with the distribution presented in \citet{kimball11}. To investigate this further requires more information on those sources below our detection limit. 

\subsection{The limitations of stacking}

One of the most obvious ways to gain this extra information on the undetected sources is to stack all of the non detections at the position of the QSO to obtain an average flux density. To ensure that the noise is approximately the same in each of the individual observations, only objects in the deep sample were stacked. This resulted in a 16$\sigma$ detection of 0.074$\pm0.005$\,mJy. 

However, rather than being an indication of the underlying flux density of the sample, this stacked emission is strongly biased towards those sources just below our detection limit. As such, this stacking detection is more indicative of our survey flux density limit rather than being an intrinsic property of the undetected QSOs. In addition, due to the large range in redshifts combined into the one image, further complications arise if we try to convert this measured average flux density into a luminosity. The large range in redshift also means that each image is weighted differently and simply taking the average or median luminosity distance will not have the same weighting.

The very nature of stacking means that we lose all information on the underlying distribution of the non detected sources and therefore it is not trivial to compare the stacked detection with the luminosity distribution presented in \citet{kimball11}. 

\begin{figure}
  \centerline{\plotone{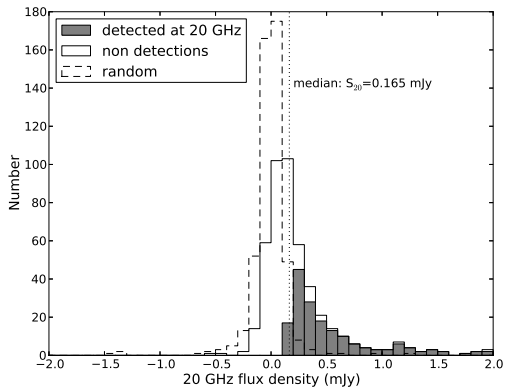}}
  \caption{Statistical probability distribution for our X-ray selected sample (solid line) which shows that while many sources are not detected in the individual images the average 20\,GHz flux density of the sample is non-zero. The dashed line shows the distribution calculated from random points which confirms that the QSO distribution is real, and the dotted line denotes the median flux density of 0.165\,mJy. Only sources with a peak flux density less than 2\,mJy are shown here, but the tail of the distribution extends to brighter flux densities.}
\label{probdist}
\end{figure}

An alternative method is to plot the flux density probability distribution function (hereafter termed flux density PDF) as shown in Figure \ref{probdist}. This is derived by measuring the flux densities at the optical positions of the targets, shown by the solid line. Sources with more than a 3$\sigma$ detection at 20\,GHz are shaded in grey. We have again only included sources in the deep sample as these images have approximately uniform noise properties. Although the non-detections do not have a significant flux density measurement in the individual images, this flux density PDF shows that the average flux density of the undetected QSOs is non-zero\footnote{When carrying out the radio observations, the optical position of the QSO was slightly offset from the phase center, so this non-zero peak in the distribution is not due to a DC offset at the phase center.}. 

To confirm that this is not a result of the noise in the images, we have also calculated the distribution taken from random points in the radio images, shown by the dashed line. The median flux density of the PDF for our X-ray selected QSOs is 0.165\,mJy, compared to the median of -0.001\,mJy for the random distribution. The median flux density of the undetected sources is 0.074\,mJy, identical to the flux density obtained from the stacked image. 

While it is again non-trivial to convert this distribution into luminosity space, this does provide information on the underlying distribution as opposed to simply obtaining an average flux density. With some sophisticated modelling it would be possible to take the QSO luminosity function presented in \citet{kimball11} and predict what would be observed at 20\,GHz by assuming either a flat or steep spectrum population and compare to the results presented here. However, this is beyond the scope of this paper and will be addressed in future works.

\subsection{Star formation rates}

Although we cannot distinguish whether the radio emission is dominated by AGN or star formation processes in the data presented here, we can calculate rough estimates of the star formation rate (SFR) expected using the following relation: 

\begin{equation}
SFR(L_{1.4})=\frac{L_{1.4}}{8.85\times10^{20}}
\end{equation}
where the SFR is in \mdot\,yr$^{-1}$ and $L_{1.4}$ is the 1.4\,GHz radio luminosity in W/Hz. This relation assumes a constant burst of star-formation over 100\,Myr \citep{sullivan01}.

Assuming a steep spectral index of $\alpha=-0.7$ and using a 20\,GHz flux density of 0.074\,mJy (the average flux density of the undetected sources), we can extrapolate a 1.4\,GHz flux density of 0.48\,mJy. At a redshift of $z=0.1$ this gives a rough SFR of $\sim10$\mdot\,yr$^{-1}$ and at $z=1$ this would be $\sim2300$\mdot\,yr$^{-1}$. Note that these estimates assume that all of the radio emission is due to star formation processes and so only represent upper limits of the star formation rates. 


\subsection{QSO evolution models}

In current galaxy evolution models, QSO activity is attributed to the accretion of cold-gas \citep{croton06,bower06}. As such, it is reasonable to assume that this same reservoir of gas also contributes to star formation. Using Herschel data, \citet{bonfield11} recently found that the star formation (calculated using the Infra-Red (IR) luminosity) is correlated with the QSO accretion luminosity, indicating that both processes are supplied by the same cold gas. This agrees with CO observations of \citet{scoville03,bertram07} who find that the majority of low luminosity QSOs reside in gas-rich galaxies. In addition, recent results of \citet{2011arXiv1110.0982P} find SFRs ranging from 3--29\,\mdot\,yr$^{-1}$ for a sample of X-ray selected AGN (not necessarily all QSOs) providing further evidence that star-formation processes play an important role at lower radio luminosities. 

According to the current picture of galaxy formation through hierarchical merging, gas-rich galaxies merge to form Ultra-Luminous Infra-Red Galaxies (ULIRGs). These mergers provide the large reservoirs of cold gas needed to fuel the AGN (and star-formation). At this early stage the galaxy is shrouded in dust, hence is very luminous in the IR. As the material begins to settle towards the center of the galaxy, AGN activity can be switched on, resulting in a QSO \citep{sanders88,Canalizo+Stockton01}. As such, it is not unreasonable to expect star formation processes to start to play a dominate role in the radio emission observed at lower luminosities. In fact, the star formation rate could also provide insight into the evolutionary stage of the QSO. During the ULIRG phase high star formation rates are expected, while lower star formation rates could indicate that the gas reservoir is close to being depleted. 

\subsection{Future Observations}

From the radio data presented here alone we cannot distinguish whether the radio emission in the low-luminosity QSOs comes from AGN or star-forming processes. However, there is mounting evidence that star formation in the QSO host becomes the dominant source of radio emission at these luminosities \citep{kimball11,padovani11,bonfield11}.

To properly distinguish whether the emission is AGN or star formation dominated, better spatial information is needed, ideally using Very Long Baseline Interferometry (VLBI). However, it is most likely that both processes contribute to the observed flux density at some level. As such, molecular line observations with the Atacama Large Millimeter Array (ALMA) can study the distribution of cold gas, potentially disentangling the gas being accreted onto the central black hole from gas responsible for star formation in the host galaxy. 

\section{Conclusions} \label{conclusions}

We have observed a sample of 874 X-ray selected QSOs that were observed as part of the 6dF Galaxy Survey at 20\,GHz to determine if the high-frequency radio luminosity distribution is bimodal. High-frequency observations provide a unique dataset where the emission is dominated by the core of the AGN, hence only showing the most recent activity. Deep observations at 20\,GHz show no evidence for a bimodal distribution in either the radio luminosity distribution or in the distribution of the R$_{20}$ parameter (the ratio of optical to radio luminosities). The broad range of radio luminosities observed for a very narrow range in optical luminosity indicates that the accretion power (approximated by the optical luminosity) can result in a very wide range of jet luminosities. 

Investigating the redshift effects of these distributions clearly show a strong redshift-luminosity degeneracy, which if not taken into account, could lead to an apparent bimodal distribution. This effect could explain the contradictory results seen in previous studies. 

Recent results have suggested that at low luminosities star formation can dominate the radio emission observed in QSOs. To investigate this further we attempted to stack the undetected sources at 20\,GHz. This results in a 16$\sigma$ detection of 0.074$\pm0.005$\,mJy, but is heavily biased towards those sources just below the detection threshold. Instead we derive the flux density probability distribution by calculating the peak flux density at the position of all sources in the sample, which has a median flux density of 0.165\,mJy. This provides more information on the underlying distribution than is achieved by stacking. By modelling the QSO luminosity functions presented in \citet{kimball11} it would be possible to predict the 20\,GHz luminosity function assuming either a flat or steep spectrum population and compare with the results presented in this paper. 

While this modelling is beyond the scope of this paper, we can obtain rough estimates of the star formation rate assuming all of the radio emission in the low luminostiy QSOs is due to star formation. Using the average 20\,GHz flux density calculated for the undetected sources, this corresponds to star formation rates of $\sim10$\mdot\,yr$^{-1}$ at $z=0.1$ to $\sim2300$\mdot\,yr$^{-1}$ at $z=1$.



\acknowledgments

EKM would like to thank all of the staff at the Australia Telescope Compact Array in Narrabri, N.S.W. for their help and support during the observations. We also thank Paul Hancock for useful discussions and the referee for suggestions that greatly improved the paper. The Australia Telescope Compact Array is part of the Australia Telescope which is funded by the Commonwealth of Australia for operation as a National Facility managed by CSIRO. We acknowledge the support of the Australian Research Council through the award of an ARC Australian Professorial Fellowship (DP0451395) to EMS and an ARC QEII Fellowship (DP0666615) to SMC.







\bibliographystyle{apj}

\bibliography{bibliog}

\begin{thebibliography}{45}
\expandafter\ifx\csname natexlab\endcsname\relax\def\natexlab#1{#1}\fi

\bibitem[{{Becker} {et~al.}(1995){Becker}, {White}, \& {Helfand}}]{first}
{Becker}, R.~H., {White}, R.~L., \& {Helfand}, D.~J. 1995, \apj, 450, 559

\bibitem[{{Bertram} {et~al.}(2007){Bertram}, {Eckart}, {Fischer}, {Zuther},
  {Straubmeier}, {Wisotzki}, \& {Krips}}]{bertram07}
{Bertram}, T., {Eckart}, A., {Fischer}, S., {Zuther}, J., {Straubmeier}, C.,
  {Wisotzki}, L., \& {Krips}, M. 2007, \aap, 470, 571

\bibitem[{{Best} {et~al.}(2005){Best}, {Kauffmann}, {Heckman}, {Brinchmann},
  {Charlot}, {Ivezi{\'c}}, \& {White}}]{best05}
{Best}, P.~N., {Kauffmann}, G., {Heckman}, T.~M., {Brinchmann}, J., {Charlot},
  S., {Ivezi{\'c}}, {\v Z}., \& {White}, S.~D.~M. 2005, \mnras, 362, 25

\bibitem[{{Bonfield} {et~al.}(2011){Bonfield}, {Jarvis}, {Hardcastle},
  {Cooray}, {Hatziminaoglou}, {Ivison}, {Page}, {Stevens}, {de Zotti}, {Auld},
  {Baes}, {Buttiglione}, {Cava}, {Dariush}, {Dunlop}, {Dunne}, {Dye}, {Eales},
  {Fritz}, {Hopwood}, {Ibar}, {Maddox}, {Micha{\l}owski}, {Pascale}, {Pohlen},
  {Rigby}, {Rodighiero}, {Serjeant}, {Smith}, {Temi}, \& {van der
  Werf}}]{bonfield11}
{Bonfield}, D.~G., {et~al.} 2011, \mnras, 416, 13

\bibitem[{{Bower} {et~al.}(2006){Bower}, {Benson}, {Malbon}, {Helly}, {Frenk},
  {Baugh}, {Cole}, \& {Lacey}}]{bower06}
{Bower}, R.~G., {Benson}, A.~J., {Malbon}, R., {Helly}, J.~C., {Frenk}, C.~S.,
  {Baugh}, C.~M., {Cole}, S., \& {Lacey}, C.~G. 2006, \mnras, 370, 645

\bibitem[{{Broderick} \& {Fender}(2011)}]{broderick11}
{Broderick}, J.~W., \& {Fender}, R.~P. 2011, \mnras, 1475

\bibitem[{{Canalizo} \& {Stockton}(2001)}]{Canalizo+Stockton01}
{Canalizo}, G., \& {Stockton}, A. 2001, \apj, 555, 719

\bibitem[{{Cirasuolo} {et~al.}(2003){Cirasuolo}, {Magliocchetti}, {Celotti}, \&
  {Danese}}]{cirasuolo03}
{Cirasuolo}, M., {Magliocchetti}, M., {Celotti}, A., \& {Danese}, L. 2003,
  \mnras, 341, 993

\bibitem[{{Condon} {et~al.}(1998){Condon}, {Cotton}, {Greisen}, {Yin},
  {Perley}, {Taylor}, \& {Broderick}}]{nvss}
{Condon}, J.~J., {Cotton}, W.~D., {Greisen}, E.~W., {Yin}, Q.~F., {Perley},
  R.~A., {Taylor}, G.~B., \& {Broderick}, J.~J. 1998, \aj, 115, 1693

\bibitem[{{Condon} {et~al.}(1980){Condon}, {Odell}, {Puschell}, \&
  {Stein}}]{condon80}
{Condon}, J.~J., {Odell}, S.~L., {Puschell}, J.~J., \& {Stein}, W.~A. 1980,
  \nat, 283, 357

\bibitem[{{Croton} {et~al.}(2006){Croton}, {Springel}, {White}, {De Lucia},
  {Frenk}, {Gao}, {Jenkins}, {Kauffmann}, {Navarro}, \& {Yoshida}}]{croton06}
{Croton}, D.~J., {et~al.} 2006, \mnras, 365, 11

\bibitem[{{Goldschmidt} {et~al.}(1999){Goldschmidt}, {Kukula}, {Miller}, \&
  {Dunlop}}]{goldschmidt99}
{Goldschmidt}, P., {Kukula}, M.~J., {Miller}, L., \& {Dunlop}, J.~S. 1999,
  \apj, 511, 612

\bibitem[{{Hooper} {et~al.}(1995){Hooper}, {Impey}, {Foltz}, \&
  {Hewett}}]{hooper95}
{Hooper}, E.~J., {Impey}, C.~D., {Foltz}, C.~B., \& {Hewett}, P.~C. 1995, \apj,
  445, 62

\bibitem[{{Ivezi{\'c}} {et~al.}(2002){Ivezi{\'c}}, {Menou}, {Knapp}, {Strauss},
  {Lupton}, {Vanden Berk}, {Richards}, {Tremonti}, {Weinstein}, {Anderson},
  {Bahcall}, {Becker}, {Bernardi}, {Blanton}, {Eisenstein}, {Fan},
  {Finkbeiner}, {Finlator}, {Frieman}, {Gunn}, {Hall}, {Kim}, {Kinkhabwala},
  {Narayanan}, {Rockosi}, {Schlegel}, {Schneider}, {Strateva}, {SubbaRao},
  {Thakar}, {Voges}, {White}, {Yanny}, {Brinkmann}, {Doi}, {Fukugita},
  {Hennessy}, {Munn}, {Nichol}, \& {York}}]{ivezic02}
{Ivezi{\'c}}, {\v Z}., {et~al.} 2002, \aj, 124, 2364

\bibitem[{{Ivezi{\'c}} {et~al.}(2004){Ivezi{\'c}}, {Richards}, {Hall},
  {Lupton}, {Jagoda}, {Knapp}, {Gunn}, {Strauss}, {Schlegel}, {Steinhardt}, \&
  {Siverd}}]{ivezic04}
{Ivezi{\'c}}, Z., {et~al.} 2004, in Astronomical Society of the Pacific
  Conference Series, Vol. 311, AGN Physics with the Sloan Digital Sky Survey,
  ed. {G.~T.~Richards \& P.~B.~Hall}, 347

\bibitem[{{Jiang} {et~al.}(2007){Jiang}, {Fan}, {Ivezi{\'c}}, {Richards},
  {Schneider}, {Strauss}, \& {Kelly}}]{jiang07}
{Jiang}, L., {Fan}, X., {Ivezi{\'c}}, {\v Z}., {Richards}, G.~T., {Schneider},
  D.~P., {Strauss}, M.~A., \& {Kelly}, B.~C. 2007, \apj, 656, 680

\bibitem[{{Jones} {et~al.}(2004){Jones}, {Saunders}, {Colless}, {Read},
  {Parker}, {Watson}, {Campbell}, {Burkey}, {Mauch}, {Moore}, {Hartley},
  {Cass}, {James}, {Russell}, {Fiegert}, {Dawe}, {Huchra}, {Jarrett}, {Lahav},
  {Lucey}, {Mamon}, {Proust}, {Sadler}, \& {Wakamatsu}}]{6df2004}
{Jones}, D.~H., {et~al.} 2004, \mnras, 355, 747

\bibitem[{{Kellermann} {et~al.}(1989){Kellermann}, {Sramek}, {Schmidt},
  {Shaffer}, \& {Green}}]{kellermann89}
{Kellermann}, K.~I., {Sramek}, R., {Schmidt}, M., {Shaffer}, D.~B., \& {Green},
  R. 1989, \aj, 98, 1195

\bibitem[{{Kimball} {et~al.}(2011){Kimball}, {Kellermann}, {Condon}, {Ivezic},
  \& {Perley}}]{kimball11}
{Kimball}, A.~E., {Kellermann}, K.~I., {Condon}, J.~J., {Ivezic}, Z., \&
  {Perley}, R.~A. 2011, ArXiv e-prints: 1107.3551

\bibitem[{{Larson} {et~al.}(2011){Larson}, {Dunkley}, {Hinshaw}, {Komatsu},
  {Nolta}, {Bennett}, {Gold}, {Halpern}, {Hill}, {Jarosik}, {Kogut}, {Limon},
  {Meyer}, {Odegard}, {Page}, {Smith}, {Spergel}, {Tucker}, {Weiland},
  {Wollack}, \& {Wright}}]{wmap7}
{Larson}, D., {et~al.} 2011, \apjs, 192, 16

\bibitem[{{Mahony} {et~al.}(2010){Mahony}, {Croom}, {Boyle}, {Edge}, {Mauch},
  \& {Sadler}}]{rass6df}
{Mahony}, E.~K., {Croom}, S.~M., {Boyle}, B.~J., {Edge}, A.~C., {Mauch}, T., \&
  {Sadler}, E.~M. 2010, \mnras, 401, 1151

\bibitem[{{Massardi} {et~al.}(2011){Massardi}, {Ekers}, {Murphy}, {Mahony},
  {Hancock}, {Chhetri}, {de Zotti}, {Sadler}, {Burke-Spolaor}, {Calabretta},
  {Edwards}, {Ekers}, {Jackson}, {Kesteven}, {Newton-McGee}, {Phillips},
  {Ricci}, {Roberts}, {Sault}, {Staveley-Smith}, {Subrahmanyan}, {Walker}, \&
  {Wilson}}]{at20ganalysis}
{Massardi}, M., {et~al.} 2011, \mnras, 412, 318

\bibitem[{{Mauch} \& {Sadler}(2007)}]{mauch07}
{Mauch}, T., \& {Sadler}, E.~M. 2007, \mnras, 375, 931

\bibitem[{{Miller} {et~al.}(1990){Miller}, {Peacock}, \& {Mead}}]{mpm90}
{Miller}, L., {Peacock}, J.~A., \& {Mead}, A.~R.~G. 1990, \mnras, 244, 207

\bibitem[{{Murphy} {et~al.}(2010){Murphy}, {Sadler}, {Ekers}, {Massardi},
  {Hancock}, {Mahony}, {Ricci}, {Burke-Spolaor}, {Calabretta}, {Chhetri}, {de
  Zotti}, {Edwards}, {Ekers}, {Jackson}, {Kesteven}, {Lindley}, {Newton-McGee},
  {Phillips}, {Roberts}, {Sault}, {Staveley-Smith}, {Subrahmanyan}, {Walker},
  \& {Wilson}}]{at20g}
{Murphy}, T., {et~al.} 2010, \mnras, 402, 2403

\bibitem[{{Padovani}(1993)}]{padovani93}
{Padovani}, P. 1993, \mnras, 263, 461

\bibitem[{{Padovani} {et~al.}(2011){Padovani}, {Miller}, {Kellermann},
  {Mainieri}, {Rosati}, \& {Tozzi}}]{padovani11}
{Padovani}, P., {Miller}, N., {Kellermann}, K.~I., {Mainieri}, V., {Rosati},
  P., \& {Tozzi}, P. 2011, \apj, 740, 20

\bibitem[{{Perley} {et~al.}(2011){Perley}, {Chandler}, {Butler}, \&
  {Wrobel}}]{evla}
{Perley}, R.~A., {Chandler}, C.~J., {Butler}, B.~J., \& {Wrobel}, J.~M. 2011,
  ArXiv e-prints: 1106.0532

\bibitem[{{Pierce} {et~al.}(2011){Pierce}, {Ballantyne}, \&
  {Ivison}}]{2011arXiv1110.0982P}
{Pierce}, C.~M., {Ballantyne}, D.~R., \& {Ivison}, R.~J. 2011, ArXiv e-prints

\bibitem[{{Rafter} {et~al.}(2009){Rafter}, {Crenshaw}, \& {Wiita}}]{rafter09}
{Rafter}, S.~E., {Crenshaw}, D.~M., \& {Wiita}, P.~J. 2009, \aj, 137, 42

\bibitem[{{Sanders} {et~al.}(1988){Sanders}, {Soifer}, {Elias}, {Madore},
  {Matthews}, {Neugebauer}, \& {Scoville}}]{sanders88}
{Sanders}, D.~B., {Soifer}, B.~T., {Elias}, J.~H., {Madore}, B.~F., {Matthews},
  K., {Neugebauer}, G., \& {Scoville}, N.~Z. 1988, \apj, 325, 74

\bibitem[{{Sault} {et~al.}(1995){Sault}, {Teuben}, \& {Wright}}]{miriad}
{Sault}, R.~J., {Teuben}, P.~J., \& {Wright}, M.~C.~H. 1995, in Astronomical
  Society of the Pacific Conference Series, Vol.~77, Astronomical Data Analysis
  Software and Systems IV, ed. {R.~A.~Shaw, H.~E.~Payne, \& J.~J.~E.~Hayes},
  433

\bibitem[{{Schmidt} \& {Green}(1983)}]{1983ApJ...269..352S}
{Schmidt}, M., \& {Green}, R.~F. 1983, \apj, 269, 352

\bibitem[{{Scoville} {et~al.}(2003){Scoville}, {Frayer}, {Schinnerer}, \&
  {Christopher}}]{scoville03}
{Scoville}, N.~Z., {Frayer}, D.~T., {Schinnerer}, E., \& {Christopher}, M.
  2003, \apjl, 585, L105

\bibitem[{{Shankar} {et~al.}(2010){Shankar}, {Sivakoff}, {Vestergaard}, \&
  {Dai}}]{shankar10}
{Shankar}, F., {Sivakoff}, G.~R., {Vestergaard}, M., \& {Dai}, X. 2010, \mnras,
  401, 1869

\bibitem[{{Sikora} {et~al.}(2007){Sikora}, {Stawarz}, \& {Lasota}}]{sikora07}
{Sikora}, M., {Stawarz}, {\L}., \& {Lasota}, J.-P. 2007, \apj, 658, 815

\bibitem[{{Stawarz} {et~al.}(2008){Stawarz}, {Sikora}, \& {Lasota}}]{stawarz08}
{Stawarz}, L., {Sikora}, M., \& {Lasota}, J.-P. 2008, in Astronomical Society
  of the Pacific Conference Series, Vol. 386, Extragalactic Jets: Theory and
  Observation from Radio to Gamma Ray, ed. {T.~A.~Rector \& D.~S.~De Young},
  169

\bibitem[{{Stocke} {et~al.}(1992){Stocke}, {Morris}, {Weymann}, \&
  {Foltz}}]{stocke92}
{Stocke}, J.~T., {Morris}, S.~L., {Weymann}, R.~J., \& {Foltz}, C.~B. 1992,
  \apj, 396, 487

\bibitem[{{Strittmatter} {et~al.}(1980){Strittmatter}, {Hill}, {Pauliny-Toth},
  {Steppe}, \& {Witzel}}]{strittmatter80}
{Strittmatter}, P.~A., {Hill}, P., {Pauliny-Toth}, I.~I.~K., {Steppe}, H., \&
  {Witzel}, A. 1980, \aap, 88, L12

\bibitem[{{Sullivan} {et~al.}(2001){Sullivan}, {Mobasher}, {Chan}, {Cram},
  {Ellis}, {Treyer}, \& {Hopkins}}]{sullivan01}
{Sullivan}, M., {Mobasher}, B., {Chan}, B., {Cram}, L., {Ellis}, R., {Treyer},
  M., \& {Hopkins}, A. 2001, \apj, 558, 72

\bibitem[{{Voges} {et~al.}(1999){Voges}, {Aschenbach}, {Boller},
  {Br{\"a}uninger}, {Briel}, {Burkert}, {Dennerl}, {Englhauser}, {Gruber},
  {Haberl}, {Hartner}, {Hasinger}, {K{\"u}rster}, {Pfeffermann}, {Pietsch},
  {Predehl}, {Rosso}, {Schmitt}, {Tr{\"u}mper}, \& {Zimmermann}}]{rass}
{Voges}, W., {et~al.} 1999, \aap, 349, 389

\bibitem[{{White} {et~al.}(2007){White}, {Helfand}, {Becker}, {Glikman}, \& {de
  Vries}}]{white07}
{White}, R.~L., {Helfand}, D.~J., {Becker}, R.~H., {Glikman}, E., \& {de
  Vries}, W. 2007, \apj, 654, 99

\bibitem[{{White} {et~al.}(2000){White}, {Becker}, {Gregg},
  {Laurent-Muehleisen}, {Brotherton}, {Impey}, {Petry}, {Foltz}, {Chaffee},
  {Richards}, {Oegerle}, {Helfand}, {McMahon}, \& {Cabanela}}]{white00}
{White}, R.~L., {et~al.} 2000, \apjs, 126, 133

\bibitem[{{Wilson} {et~al.}(2011){Wilson}, {Ferris}, {Axtens}, {Brown},
  {Davis}, {Hampson}, {Leach}, {Roberts}, {Saunders}, {Koribalski}, {Caswell},
  {Lenc}, {Stevens}, {Voronkov}, {Wieringa}, {Brooks}, {Edwards}, {Ekers},
  {Emonts}, {Hindson}, {Johnston}, {Maddison}, {Mahony}, {Malu}, {Massardi},
  {Mao}, {McConnell}, {Norris}, {Schnitzeler}, {Subrahmanyan}, {Urquhart},
  {Thompson}, \& {Wark}}]{cabb}
{Wilson}, W.~E., {et~al.} 2011, \mnras, 1230

\bibitem[{{York} {et~al.}(2000){York}, {Adelman}, {Anderson}, {Anderson},
  {Annis}, {Bahcall}, {Bakken}, {Barkhouser}, {Bastian}, {Berman}, {Boroski},
  {Bracker}, {Briegel}, {Briggs}, {Brinkmann}, {Brunner}, {Burles}, {Carey},
  {Carr}, {Castander}, {Chen}, {Colestock}, {Connolly}, {Crocker}, {Csabai},
  {Czarapata}, {Davis}, {Doi}, {Dombeck}, {Eisenstein}, {Ellman}, {Elms},
  {Evans}, {Fan}, {Federwitz}, {Fiscelli}, {Friedman}, {Frieman}, {Fukugita},
  {Gillespie}, {Gunn}, {Gurbani}, {de Haas}, {Haldeman}, {Harris}, {Hayes},
  {Heckman}, {Hennessy}, {Hindsley}, {Holm}, {Holmgren}, {Huang}, {Hull},
  {Husby}, {Ichikawa}, {Ichikawa}, {Ivezi{\'c}}, {Kent}, {Kim}, {Kinney},
  {Klaene}, {Kleinman}, {Kleinman}, {Knapp}, {Korienek}, {Kron}, {Kunszt},
  {Lamb}, {Lee}, {Leger}, {Limmongkol}, {Lindenmeyer}, {Long}, {Loomis},
  {Loveday}, {Lucinio}, {Lupton}, {MacKinnon}, {Mannery}, {Mantsch}, {Margon},
  {McGehee}, {McKay}, {Meiksin}, {Merelli}, {Monet}, {Munn}, {Narayanan},
  {Nash}, {Neilsen}, {Neswold}, {Newberg}, {Nichol}, {Nicinski}, {Nonino},
  {Okada}, {Okamura}, {Ostriker}, {Owen}, {Pauls}, {Peoples}, {Peterson},
  {Petravick}, {Pier}, {Pope}, {Pordes}, {Prosapio}, {Rechenmacher}, {Quinn},
  {Richards}, {Richmond}, {Rivetta}, {Rockosi}, {Ruthmansdorfer}, {Sandford},
  {Schlegel}, {Schneider}, {Sekiguchi}, {Sergey}, {Shimasaku}, {Siegmund},
  {Smee}, {Smith}, {Snedden}, {Stone}, {Stoughton}, {Strauss}, {Stubbs},
  {SubbaRao}, {Szalay}, {Szapudi}, {Szokoly}, {Thakar}, {Tremonti}, {Tucker},
  {Uomoto}, {Vanden Berk}, {Vogeley}, {Waddell}, {Wang}, {Watanabe},
  {Weinberg}, {Yanny}, \& {Yasuda}}]{sdss}
{York}, D.~G., {et~al.} 2000, \aj, 120, 1579

\end{thebibliography}

\end{document}